# Toward Trusted Onboard Artificial Intelligence (AI): Advancing Small Satellite Operations using Reinforcement Learning

Cannon Whitney
University of Florida
Gainesville, FL 32611; 352-392-3261
cannon.whitney@ufl.edu

Joseph Melville
Air Force Research Laboratory - Space Vehicles Directorate's Small Satellite Portfolio
Kirtland Air Force Base, NM 87117; 801-472-2196
joseph.melville.3@spaceforce.mil

## ABSTRACT

A Reinforcement Learning (RL) algorithm was developed for command automation onboard a 3U CubeSat. This effort focused on the implementation of macro Control Action Reinforcement Learning (CARL), a technique in which an onboard agent is provided with compiled information based on live telemetry as its observation. The agent uses this information to produce high-level actions, such as adjusting attitude to solar pointing, which are then translated into control algorithms and executed through lower-level instructions. Once trust in the onboard agent is established, real-time environmental information can be leveraged for faster response times and reduced reliance on ground control. The approach not only focuses on developing an RL algorithm for a specific satellite but also sets a precedent for integrating trusted AI into onboard systems. This research builds on previous work in three areas: (1) RL algorithms for issuing high-level commands that are translated into low-level executable instructions; (2) the deployment of AI inference models interfaced with live operational systems, particularly onboard spacecraft; and (3) strategies for building trust in AI systems, especially for remote and autonomous applications. Existing RL research for satellite control is largely limited to simulation-based experiments; in this work, these techniques are tailored by constructing a digital twin of a specific spacecraft and training the RL agent to issue macro actions in this simulated environment. The policy of the trained agent is copied to an isolated environment, where it is fed compiled information about the satellite to make inference predictions, thereby demonstrating the RL algorithm's validity on orbit without granting it command authority. This process enables safe comparison of the algorithm's predictions against actual satellite behavior and ensures operation within expected parameters.

## INTRODUCTION

Satellite missions and operations are experiencing a significant increase in complexity, driven by the proliferation of small satellites. This growth has driven demand for advanced Earth observation capabilities and deep-space exploration objectives.[1] Traditional ground-based satellite management is now under strain, struggling to keep pace with the dynamic interactions of modern missions.[2] As satellite constellations expand, the volume and repetitiveness of operational tasks become prohibitively burdensome.[3] Even with automated scripts in place, human-operated systems remain limited by operator situational awareness and reaction time, which constrains overall responsiveness.[4] This growing gap between established operational methodologies and the evolving requirements of advanced satellite missions underscores the need for new approaches to effectively manage the future space ecosystem.

Despite these growing needs, most modern satellite missions still rely primarily on procedural automation that requires significant human oversight[5]. These pre-scripted command sequences, while effective for basic task execution, fall short of the full autonomy needed for modern mission demands.

Stanford University's Space Rendezvous Laboratory (SLAB) conducted the first on-orbit demonstration of autonomous swarm navigation using the Starling Formation-Flying Optical Experiment (StarFOX).[6] In this mission, a group of four CubeSats maintained relative positioning through onboard visual sensing and inter-satellite data exchange, operating independently of GPS or ground-based guidance systems.[6] While this effort is a significant step toward trusted automation in space, the Absolute and Rela-tive Trajectory Measurement System (ARTMS) supporting StarFOX achieves this autonomy using classical estimation and tracking algorithms rather than any machine learning technique.[6]

Although non-neural network-based approaches can achieve a certain degree of autonomy, there is a threshold beyond which further advancements require substantial manual effort, making neural network-based methods necessary. Neural networks can outperform traditional rule-based systems in complex control tasks due to their multi-layered architectures, which efficiently capture subtle, nonlinear relationships between inputs and outputs that would otherwise be impractical to represent.[7] This research aims to explore methods for establishing trust in machine learning-based systems to enhance performance beyond even the capabilities of StarFOX.

Although many recent efforts to enable greater onboard autonomy are documented widely in the literature, most have not progressed beyond virtual testing. Specifically, research addressing autonomous satellite tasking through any type of RL has predominantly emphasized simulation-based experiments, with little validation under operational conditions. Each of the following studies presents simulated validation results, with some also proposing operational models or more advanced scenarios. The RL-based scheduling described by D. A. Zeleke and H.-D. Kim[8] outperformed the heuristic method by scheduling over 85% of feasible imaging tasks compared to just 50% with heuristics, while also utilizing resources more efficiently. V. Bajenaru, A. Herrmann, H. Schaub, J. Ramirez, and S. Phillips[9] developed a Deep Reinforcement Learning (DRL) policy which outperformed the rule-based policy by fully avoiding activations of the Run Time Assurance safety system (0% vs. 6.85%) and achieving slightly higher imaging success, with 34.09 versus 33.77 targets imaged.

A. Kyuroson, A. Banerjee, N. A. Tafanidis, S. Satpute, and G. Nikolakopoulos[10] implemented DRL algorithms that were successful in autonomously controlling low-Earth orbit satellites to maintain precise station-keeping within tight tolerance margins, despite external perturbations and realistic low-thrust constraints. The agent developed by K. Naik, O. Chang, and C. Kotulak[11] recovered from every cloud occlusion event within 1 minute by retasking or reorienting the satellite, thereby converting otherwise unproductive collection periods into useful operations. H. H. Lei, M. Shubert, N. Damron, K. Lang, and S. Phillips[12] use hierarchical multiagent DRL to collaboratively inspect all required viewpoints of a defunct satellite by combining independently trained guidance and navigation policies. These successes have yielded a growing inventory of fully trained RL control models, but to date, these models have only been validated in simulation-based environments.

Earning operator trust in deep learning tools for application beyond simulated environments is difficult, and requires even more than testing on increasingly representative hardware and environments. The remote and inaccessible nature of satellites amplifies concerns regarding the reliability of autonomous systems, especially given that neural network-based algorithms often function as "black boxes".[13] Therefore, beyond merely testing in representative environments, additional efforts are needed to improve operator understanding of deep learning tools. Towards this end, there are established methods for interpreting the internal logic of these systems.[13] A standard approach to elucidate the behavior of a "black box" algorithm is to thoroughly sample its input space and analyze the corresponding outputs. This allows the model's behavior to be better understood prior to use; R. Guidotti, A. Monreale, S. Ruggieri, F. Turini, F. Giannotti, and D. Pedreschi[13] refers to this general process as reverse engineering the black box.

As an application of this approach, the deployment of autonomous agents for a satellite system can follow a staged approach that gradually builds operator confidence while verifying model behavior under realistic conditions. After training the agent in simulation, one of the first steps is to test it on historical telemetry from a comparable satellite. By feeding in past mission data, human operators can observe the agent's responses and assess whether it behaves properly, within their estimation. As a sanity check, the model can be placed in simple high-stakes situations, such as a critically low power state while oriented away from the sun, to confirm it takes intuitive actions like entering a solar-pointing mode. This process aligns with reverse engineering the black box: gaining insights into the agent's decision logic before deployment.[13] This type of sampling supports engineers and operators gaining a clear understanding of the agent's behavior under defined conditions.

A partnership between researchers at Saint Louis University and Bennett Research Technologies has proposed a mission called Distributed Observation Reasoning and Reaction Experiment (DORRE) that directly addresses the need to build trust with operators. The DORRE mission proposes an eight-3U CubeSat constellation, beginning with the soon-to-launch DARLA pathfinder, that autonomously detects and responds to events in low Earth orbit.[14] To compare onboard reasoning with human-operated performance, the mission will conduct a series of parallel experiments in which both the autonomous agents and student human teams respond to identical event scenarios.[14] Each side's effectiveness will be evaluated across cost, timeliness, quality, and operational risk.[14] Although DORRE has not yet been flown, the mission would provide invaluable methods for building trust in autonomous tools for space system applications. However, the DORRE mission stops short of outlining a comprehensive end-to-end approach for building trust in autonomous tools, which this work seeks to rectify.

Toward this end, this work first details a phased trust-building process for maturing autonomous capabilities. These phases are then applied to develop, integrate, and deploy a DRL algorithm for command automation. A DRL application was chosen as an example because it represents one of the most complex and least inherently trustworthy autonomy tools to deploy. The phases of automation trust-building employed in this paper are contained in three categories with two phases each:

- Development phases:
  1) Traditional training, validation, and testing of the model on a digital twin
  2) Apply explainability methods that attribute model behavior to mechanisms understood by operators
- Integration phases:
  3) Test model interfaces and functions on ground engineering model
  4) Test model interfaces and functions on flight assets
- Deployment phases:
  5) Deploy as recommendation engine, where humans approve each action
  6) Deploy as a fully autonomous agent, humans pre-approve action types

Each step is outlined and phases 1 through 4 are demonstrated. The implementation of phases 5 and 6 is discussed for future work. The on-orbit asset utilized in this work is a 3U cubesat referred to as LIME, which was owned and operated by a third party. While the development of the DRL algorithm remains a work in progress, key lessons learned include:

- Implementation of a sampling method to reverse engineer the "black box" of the DRL agent, and to measure operator-relevant model performance.
- Container development and integration for on-board operations, including methods to significantly reduce the size of the original image, and increase on-orbit re-write flexibility.
- Integration of keyed and validated data structures to ensure data alignment and unit consistency.

## BACKGROUND

This section describes in further detail concepts that are common to the area of satellite control automation and foundational to the methods of this work.

Reinforcement Learning (RL) is a technique in which a software agent interacts with an environment by receiving observations in the form of an array of numbers, then calculates an optimal action to control the system to a desired state. The agent learns from feedback provided by a reward function that rates the optimality of the agent's decisions.[15] Behaviors that lead to higher cumulative rewards are reinforced and become more likely in future steps, enabling the agent to improve its decision-making over time.[15] Traditional RL methods, where deep learning is not used, are usually limited to mapping linear relationships within the input space.[15] While these methods are useful in resource-constrained environments, they often struggle to generalize to high-dimensional state and action spaces encountered in real-world satellite operations.[15] Deep Reinforcement Learning (DRL) enhances traditional RL by using deep neural networks to capture complex, nonlinear relationships in observation data, improving performance in noisy and dynamic scenarios.[15] Whether an RL system is considered deep or traditional depends on the training algorithm used, which is responsible for updating the policy based on the agent's experience.[15]

Proximal Policy Optimization (PPO) is a popular choice of training algorithm for autonomous vehicles due to its stability and relative simplicity compared to alternative DRL algorithms like deep Q-learning and trust region policy optimization.[16] This remains true for autonomous satellite systems, where PPO's stability and sample efficiency make it well-suited for handling complex, continuous control tasks.[17,18] After being trained, the resulting policy can be extracted and used for inference, a process in which the policy receives observations and generates actions based on its learned behavior, without further learning or updates.

BSK-RL is an open-source Python package that integrates RL with high-fidelity spacecraft simulations by wrapping the Basilisk astrodynamics simulation framework within the Gymnasium API.[19] This integration allows researchers to develop and test RL algorithms in realistic space mission scenarios, leveraging Basilisk's modular and scalable simulations of spacecraft dynamics, including aspects like attitude control, power systems, and environmental interactions, all accessible through a Python interface.[20] By utilizing the standardized Gymnasium API, BSK-RL ensures compatibility with major RL training libraries such as ray.[21] Compared to non-Gymnasium-based RL frameworks, BSK-RL offers significant advantages for spacecraft autonomy research, including high-fidelity simulations that reduce the gap between simulation and actual deployment, a standardized interface that promotes interoperability and ease of use, and an open-source nature with comprehensive documentation that supports reproducibility and collaborative development within the research community.[19]

## METHODOLOGY

Advancing reliable onboard autonomy for small satellites requires a cohesive framework spanning training, integration, and in-flight validation, as described in the following methodology. The methodology is structured in three main phases: development, integration, and deployment. During the development phase, a digital twin of the spacecraft environment was implemented in the BSK-RL gymnasium framework, where the agent is trained to interact with the simulated satellite environment using the PPO algorithm.[19] Development also includes the procedures used to validate the DRL model using reverse engineering. The integration phase focuses on adapting the trained policy for onboard inference by abstracting telemetry into the same input format used in simulation and logging the results. Integration also details the containerized implementation of the inference pipeline on the onboard computer. Finally, the deployment phase consists of the gradual transfer of the model from a recommendation engine to a fully autonomous agent, certified piecemeal by the operator. Each phase is described in detail in the sections that follow.

### Development:

During the development phase, the following mission-specific parameters were incorporated into the digital twin simulation: mass, volume, solar panel characteristics, reaction wheel specifications, power consumption, and planned orbital path. This design supports a modular approach, allowing adaptation for similar CubeSat missions by simply updating configuration files or parameter sets. After setup, the RL agent is trained using the PPO algorithm on the distributed parallel computing platform Ray.[21] For the initial simulation, the satellite agent is placed in a survival-focused scenario, where it is rewarded based on the proportion of time it remains operational relative to the total episode duration. This means that if the satellite successfully avoids failure states for the entire simulation, it accumulates a reward close to 1, with minor deviations due to floating-point precision. Terminal states were defined to reflect unrecoverable failures, particularly complete battery depletion or reaction wheel saturation. Entering any of these conditions would immediately end the episode and result in a negative reward less than -1, signaling undesirable behavior and guiding the agent to avoid similar situations.

Training was performed on a single desktop computer equipped with an NVIDIA 4090 GPU and an Intel i9 CPU, using a standard Ray Docker container based on an `nvidia/cuda` image to leverage GPU acceleration.[22] The containerized approach provided a reproducible environment, opening the possibility for potential future expansions to distributed training clusters by deploying multiple containers. During the training phase, checkpoints were automatically saved every five episodes, where each episode represented six months of simulated mission operations, thereby exposing the agent to diverse environmental and operational conditions. After training concluded, the policy achieving the highest average reward was selected, extracted from its checkpoint, and deployed for inference.

The observation properties summarized in Table 1 define the agent's state interface using quantities that are either directly available as telemetry or can be derived from routinely measured data. The order of the rows in Table 1 reflects the onboard calculation sequence; for example, $[BN]$ is determined first and forms the basis for deriving $\sigma_{BN}$, as well as generating values used to calculate ${}^{\mathcal{B}}\omega_{\mathrm{BN}}$, ${}^{\mathcal{N}}r_{\mathrm{BN}}$, and ${}^{\mathcal{N}}v_{\mathrm{BN}}$. These properties can be grouped into three categories: health system metrics $(z, \omega)$, attitude representations $([BN], \sigma_{BN})$, and dynamic state variables (${}^{\mathcal{B}}\omega_{\mathrm{BN}}$, ${}^{\mathcal{N}}r_{\mathrm{BN}}$, ${}^{\mathcal{N}}v_{\mathrm{BN}}$). These quantities provide the agent with insight into the spacecraft's operational state.

The macro control actions summarized in Table 2 define the agent's high-level command options for spacecraft operation. Each action–Drift, Charge, and Desaturate–is listed with a brief description and a set of Equivalent Operator Instructions. These instructions are presented as high-level pseudocode that resembles the process a human operator would follow to control the satellite in this manner. If the algorithm were given operational authority, it would issue these commands directly, and they would execute Flight Software (FSW) programs implementing the pseudocode.

**Table 1: Compiled from M. Stephenson[23]**

| Observation Property | Explanation | Telemetry Sources |
|---|---|---|
| $[BN]$ | Direction Cosine Matrix, describes the spacecraft body frame $\mathcal{B}$ with respect to the inertial frame $\mathcal{N}$. Element $BN_{ij} = \hat{b}_i \cdot \hat{n}_j$.[24] | Estimated Roll, Pitch, Yaw; X, Y, Z Position; and X, Y, Z Velocity |
| $\sigma_{BN}$ | Modified Rodrigues Parameter (MRP) vector, a 3-dimensional representation of the attitude of $\mathcal{B}$ relative to $\mathcal{N}$.[25] | Derived from $[BN]$ |
| $^{\mathcal{B}}\omega_{BN}$ | Instantaneous angular velocity of $\mathcal{B}$ relative to $\mathcal{N}$, expressed in the $\mathcal{B}$ frame.[25] | X, Y, Z Estimated Angular Rate; and Values Calculated while Generating $[BN]$: X, Y, Z Position; and X, Y, Z Velocity |
| $^{\mathcal{N}}r_{BN}$ | Position of $\mathcal{B}$ relative to $\mathcal{N}$, expressed in the $\mathcal{N}$ frame. | Values Calculated while Generating $[BN]$: X, Y, Z Position |
| $^{\mathcal{N}}v_{BN}$ | Velocity of $\mathcal{B}$ relative to $\mathcal{N}$, expressed in the $\mathcal{N}$ frame. | Values Calculated while Generating $[BN]$: X, Y, Z Velocity |
| $z$ | Battery charge fraction, current amount of stored energy divided by storage capacity. | Voltage, Current, Current Direction |
| $\omega$ | Wheel speed fraction, current reaction wheel speed divided by the maximum safe speed. | X, Y, Z Wheel Speeds |

Alternatively, if the agent's role is to provide recommendations, the suggested actions would be translated into operator instructions derived from this pseudocode. This structure supports consistent translation of agent decisions into actionable procedures, providing a clear interface for both autonomous and supervised operation.

**Table 2: Compiled from M. Stephenson[26]**

| Macro Control Action | Description | Equivalent Operator Instructions |
|---|---|---|
| Drift | Satellite cancels any queued tasks. | 1  task queue $\leftarrow \emptyset$ |
| Charge | Slews solar panels toward the sun to maximize charging. | 1  $\omega \leftarrow$ chosen rotation rate<br>2  $\sigma_{BN} \leftarrow$ determine(current attitude)<br>3  $^{\mathcal{B}}\hat{s} \leftarrow$ determine(solar angle using sun-sensor)<br>4  slew($\sigma_{BN} \rightarrow {}^{\mathcal{B}}\hat{s}$) at $\omega$ |
| Desaturate | Maneuvers the satellite to a desaturation attitude, and begins momentum unloading of the reaction wheels. | 1  $\omega \leftarrow$ chosen rotation rate<br>2  $^{\mathcal{B}}\hat{d} \leftarrow$ chosen desaturation attitude *usually nadir or solar*<br>3  $\sigma_{BN} \leftarrow$ determine(current attitude)<br>4  slew($\sigma_{BN} \rightarrow {}^{\mathcal{B}}\hat{d}$) at $\omega$<br>5  unload reaction wheel momentum using thrust system |

To support iterative validation and improvement, representative input spaces were generated and systematically sampled to test the policy's behavior in well-defined scenarios. This evaluation was visualized through diagnostic plots where each axis represented a key system state, such as reaction wheel saturation on the x-axis and battery charge percentage on the y-axis. Representative background features drawn from the simulation were used to populate the remaining input dimensions. Each point on the plot corresponds to a sampled state, with the color encoding the action selected by the policy. Several rendering methods could be used to visualize the agent's decisions; specifically, sampling from the policy's stochastic output to preserve its probabilistic nature; selecting the action with the highest logit to represent the most likely deterministic choice; or blending colors to illustrate the likelihood of each action being selected in a given state. In the latter case, each action was assigned a distinct primary color, and the resulting hues were mixed proportionally to reflect the agent's overall behavior across the action distribution. As one moves along the axes toward more extreme operational states, changes in color help illustrate how the policy's behavior shifts in response to increased environmental pressure. This data can be used to improve the simulation environment. For example, if the agent fails to act proactively when battery levels are low, parameters such as maximum battery capacity or power draw can be adjusted to increase the likelihood of terminal states resulting from insufficient charging. This method can be used to detect discrepancies or errors in the deployment pipeline by illuminating inconsistent inference output between ground and onboard systems.

### Integration:

After training, the RL agent's policy is exported and wrapped in an inference script designed for integration onboard the satellite. Rather than processing raw telemetry directly, another script derives features from the raw telemetry that replicates the abstracted input values the agent was exposed to during simulation. Table 1 describes the datapoints that are used to calculate the input values. Reducing the input space through abstraction is beneficial for RL training, as it shrinks the search space, allowing for an optimal policy to be converged on faster. This process also increases consistency between the training and operational environments, so the onboard system provides inputs in the format the policy expects, and this reduces the impact of discrepancies between simulated and real spacecraft data.

Once the input features are computed from telemetry, the onboard inference script passes them to the imported policy. The policy selects a control action, which can be decompiled into a concrete command sequence executable by the flight computer. The current version generates only the theoretical actions the agent would take if it were in control, creating a safe test environment for the integration and evaluation of autonomous technologies with no operational risk. Table 2 includes examples of how macro control actions may be expanded into FSW commands.

All of this runs within a Docker containerized environment operating on LIME's on-board computer during the mission, which hosts several other onboard computational experiments. Communication between the host satellite bus computer and the Docker container is established through open ports, allowing the container to receive telemetry data directly from the host system. In addition, directories from the host are mounted into the container, providing a mechanism for transferring files such as updated inference policy files or experiment results. While new Docker images can be uploaded to update the container, the process can be time-consuming due to the size of the container images, which were large even after compression. To support more agile updates, standalone bash scripts can also be uploaded and executed within the container, as well as the importing of new inference policy files. The advantage of this containerized approach is that it isolates experimental payloads from the core FSW, significantly reducing the risk of running novel space software.

### Deployment:

Deployment of the model describes methods by which the model can be used to iteratively build operator trust. Each iterative deployment step gives the operator to observe model performance, initially approving actions individually, and later collectively, as confidence in the autonomous system grows. In this work, these iterative phases include deployment first as a recommendation engine and eventually as a fully autonomous agent. In practice, the distinction between a recommendation engine and a fully autonomous agent will often be blurred; as operators gain trust in certain decision categories, they can selectively designate these as trusted for autonomous execution, while retaining manual oversight for less trusted aspects of the system.

The deployment phases described above were not implemented herein and remain an object of future work.

## RESULTS

This section presents outcomes from the project's development and integration categories, leaving the deployment phases to future work. These results highlight specific technical challenges encountered throughout the process and summarize the effectiveness of the implemented solutions. The structure of the results section mirrors the methodology, with dedicated subsections for each phase category, allowing for a correspondence between the experimental approach and results.

### Development

During the development phase, significant effort was made to implement magnetorquers on the digital twin within the BSK-RL framework. The default reaction wheel desaturation behavior within the Basilisk simulation relies upon thruster systems; however, the LIME satellite does not have traditional thrusters and uses magnetorquers instead. Despite multiple integration attempts, repeated errors arose when attempting to configure the FSW to enable the RL agent to use torque rods instead of thrusters. Given the project's limited timeline and recognizing that the RL agent primarily selects general actions without explicit hardware dependencies, the decision was made to proceed without full magnetorquer integration.

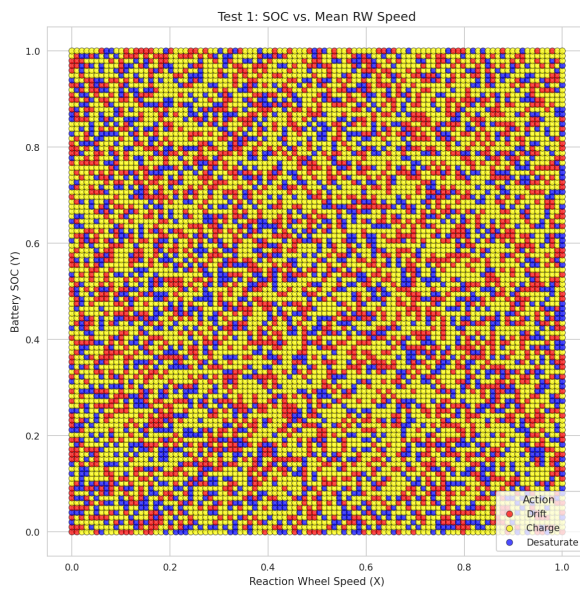

**Figure 1: Average Reaction Wheel Saturation Percent vs. Battery Charge Percent**

Additionally, challenges arose when evaluating the trained inference policy under controlled "sanity check" environments. In these scenarios, two health system metrics–such as reaction wheel speed percentage and battery charge fraction–were systematically varied and plotted on the axes, so that only data points at the extremes of the x or y axis represented clear-cut cases. All other input values, unrelated to the tested health system metrics, were set to nominal levels sampled from standard simulation runs. Initial evaluation revealed the inference policy's decisions appeared inconsistent and randomized, failing to act as expected even in the obvious cases. This performance was assessed by visualizing policy decisions graphically using a plot. Figure 1 is an example of this type of plot, with reaction wheel saturation mapped on the x-axis and battery charge on the y-axis.

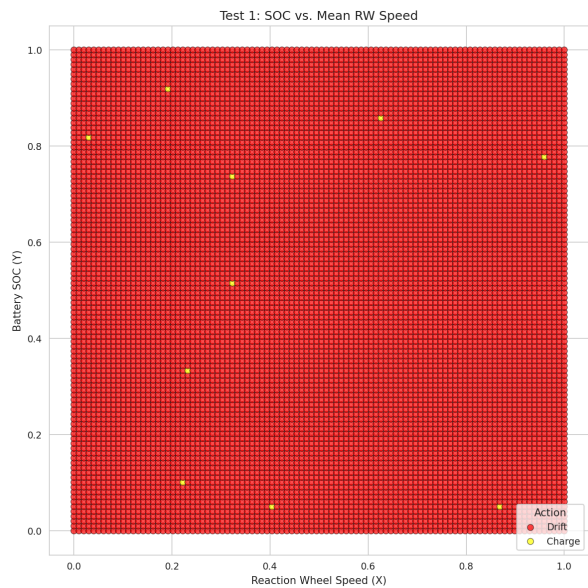

**Figure 2: Average Reaction Wheel Saturation Percent vs. Battery Charge Percent After Improved Training**

To address this inconsistency, the simulation environment's difficulty was increased by reducing battery capacity and reaction wheel maximum rotational speed thresholds. This strategy slightly improved the agent's reactiveness by forcing it to learn more proactive survival strategies. Subsequent evaluations demonstrated an improvement in performance, as depicted in Figure 2. While a pattern has emerged, the agent consistently drifts and remains unresponsive, failing to change its behavior in response to different inputs. This indicates that, even with increased environmental difficulty, the agent is not learning to adapt its actions. Ensuring the agent develops more responsive control strategies will be a focus of future work.

### Integration

Containerization initially resulted in an excessively large Docker image, posing deployment challenges. To address this, a multi-stage Docker build process was implemented using the same base image, `python:3.9-slim`, for both the builder and production stages. In the builder stage, all necessary development tools and dependencies required to compile and assemble the application were installed. After the build process was complete, only the essential runtime artifacts, such as compiled binaries and configuration files, were copied into the production stage. This approach ensured that development tools and unnecessary files were excluded from the final image, significantly reducing its size.

Collaborative communication with the third-party owner of the on-orbit satellite was essential to determine the appropriate network ports and mounted file directory structures between the host and container. Furthermore, complexities emerged from cross-compilation between the development environment's x86 architecture and the satellite bus computer's native processor. To facilitate building compatible images on x86 hosts, several developers configured QEMU-based cross-compilation capabilities. An added benefit of this approach was a further reduction in image size, as the dependencies for the satellite computer processor were smaller.

During integration with the ground engineering model, it was determined that a simple method for updating configuration files within the container was needed. This concept was implemented by directing the container to look for files of a specific name after each cycle. In this way, those files could be transferred to the specific directory on the satellite, noticed by the onboard computer, and used to update configuration files, save a new DRL model for use in future cycles, and even rewrite the functionality of the container. This allows for iterative development, even after launch of the satellite, without needing to upload an entirely new container image.

Another integration challenge involved data downlinking constraints, which became evident when the initial configuration attempted to downlink a volume of telemetry data that exceeded the available bandwidth allocation. To address this, the container was reconfigured to reduce the size of transmitted data by removing all information not directly related to control action outputs, and by applying additional data compression techniques to the remaining fields. While this approach ensured compliance with bandwidth limita-tions, it also eliminated critical contextual information, making it difficult to reliably correlate original telemetry with generated model outputs. Following a subsequent increase in the permitted data limit, an updated script was implemented to reintroduce essential telemetry, input, and observation fields, while continuing to utilize data compression to optimize transmission.

A significant post-launch challenge was structural data misalignment. Telemetry data columns compiled for the RL model were misaligned with the policy's expected structure. For example, reaction wheel speeds appeared in incorrect columns, resulting in processing errors. To address this, a Python-based translation class was introduced, using keyword mappings instead of numeric indices, greatly reducing future risk of similar mistakes.

Further investigations uncovered discrepancies in unit standards between telemetry data and simulation parameters. To permanently address these inconsistencies, ongoing refactoring using the `unyt` library was initiated, embedding unit consistency directly into the codebase.[27] Additionally, code restructuring identified significant software dependencies that, once removed, substantially reduced Docker container size. While fully developed, deployment of this optimized code remains pending due to the magnitude of updates involved. These experiences highlight critical lessons about rigorous pre-launch validation, and solutions were designed to ensure robust and reliable future iterations.

### CONCLUSIONS

This research presents a step toward integrating trusted AI into onboard small satellite systems, specifically through the development and phased validation of Macro CARL, a DRL algorithm for command automation. By employing a digital twin of a 3U CubeSat, the agent was trained to issue high-level actions based on compiled live telemetry, moving beyond traditional simulation-limited studies.

The methodology delineated a clear, phased approach to building trust in autonomous systems: beginning with development and testing in a simulated environment, followed by careful integration with ground engineering models and flight assets, and finally, a proposed gradual deployment as a recommendation engine before full autonomy. While the deployment phases are earmarked for future work, the successful demonstration of the development and integration phases lays crucial groundwork.

Key lessons learned include the necessity of refining simulation environments to elicit proactive agent behavior, as evidenced by the improved responsiveness after increasing environment difficulty. Challenges in integrating specific hardware (magnetorquers) and the initial inconsistencies in the inference policy's decisions highlighted the iterative nature of development and the importance of systematic evaluation through input space sampling and visualization. The integration phase also yielded vital insights into optimizing containerized environments for onboard operations, including significant image size reduction through multi-stage Docker builds and establishing flexible configuration update mechanisms.

This work not only contributes toward a DRL algorithm tailored for small satellite operations but also establishes a reproducible framework for developing and validating trusted onboard AI. The ability to safely compare the algorithm's predictions against actual satellite behavior on orbit, without granting direct command authority, is a critical step towards establishing operator trust and enabling real-time, autonomous responses in future space missions.

**FUTURE WORK**

Future research will follow three vectors, the first is to continue developing Macro CARL for autonomous satellite commanding. Future development will address the limitations of the current reward structure. By moving beyond the existing survival-based reward function and implementing more dynamic, mission-relevant reward schemes, the framework will facilitate the emergence of more complex and adaptive agent behavior. Such improvements are expected to enhance the agent's ability to achieve mission objectives in realistic environments. As improvements are made, the development and integration phases of the trust-building framework presented in this work will be iterated until the tool passes both traditional training and model sampling validation methods.

The second vector for future work includes the completion of deployment phases, first as a recommendation engine and gradually promoting the algorithm to full autonomy as certified by the operator. To enable this gradual promotion process, model functionality will need to be split into portions that can be promoted separately from other model functions.

The third vector for future work includes efforts to expand on the experimental framework, including systematic comparisons between satellite-specified and generic algorithms. Evaluating performance across both mission-tailored and broadly applicable agents will help quantify how closely the training environment must resemble operational conditions for the agent to consistently generate beneficial actions. These studies aim to clarify the trade-offs between digital twin fidelity and algorithm generalizability, establishing practical guidelines for deploying RL models across diverse satellite missions.

Additional work will focus on integrating a broader range of training algorithms into the modular inference system. Beyond the current implementation with PPO, alternative learning approaches will be incorporated to assess their suitability and effectiveness for onboard satellite control. This expansion will support benchmarking and cross-validation of different RL strategies under comparable operational scenarios. After establishing a baseline with these methods, the framework will be further extended to accommodate non-reinforcement learning algorithms, facilitating direct comparison across a broader set of autonomous control techniques.

*References*

[1] I. Siddique, "Small Satellites: Revolutionizing Space Exploration and Earth Observation," no. 4908526. Rochester, NY, Mar. 31, 2024. [Online]. Available: https://papers.ssrn.com/abstract=4908526

[2] W. Liu, M. Wu, G. Wan, and M. Xu, "Digital Twin of Space Environment: Development, Challenges, Applications, and Future Outlook," *Remote Sensing*, vol. 16, no. 16, p. 3023, Jan. 2024, doi: 10.3390/rs16163023.

[3] F. Baccelli, S. Candel, G. Perrin, and J.-L. Puget, "Large Satellite Constellations: Challenges and Impact," Mar. 2024. doi: 10.62686/3.

[4] M. Jipp, "Reaction Times to Consecutive Automation Failures: A Function of Working Memory and Sustained Attention," *Human Factors*, vol. 58, no. 8, pp. 1248–1261, Dec. 2016, doi: 10.1177/0018720816662374.

[5] M. K. Ben-Larbi *et al.*, "Towards the Automated Operations of Large Distributed Satellite Systems. Part 1: Review and Paradigm Shifts," *Advances in Space Research*, vol. 67, no. 11, pp. 3598–3619, Jun. 2021, doi: 10.1016/j.asr.2020.08.009.

[6] J. Kruger, S. S. Hwang, and S. D'Amico, "Starling Formation-Flying Optical Experiment: Initial Operations and Flight Results." Jun. 10, 2024. doi: 10.48550/arXiv.2406.06748.

[7] A. T. Harris, "Autonomous Management and Control of Multi-Spacecraft Operations Leveraging Atmospheric Forces," 2021. [Online]. Available: https://hanspeterschaub.info/Papers/grads/AndrewHarris.pdf

[8] D. A. Zeleke and H.-D. Kim, "A New Strategy of Satellite Autonomy with Machine Learning for Efficient Resource Utilization of a Standard Performance CubeSat," *Aerospace*, vol. 10, no. 1, p. 78, Jan. 2023, doi: 10.3390/aerospace10010078.

[9] V. Bajenaru, A. Herrmann, H. Schaub, J. Ramirez, and S. Phillips, "Trustworthy Reinforcement Learning for Decentralized Control of Satellites," Feb. 2023, [Online]. Available: https://www.researchgate.net/publication/367453124_Trustworthy_Reinforcement_Learning_for_Decentralized_Control_of_Satellites

[10] A. Kyuroson, A. Banerjee, N. A. Tafanidis, S. Satpute, and G. Nikolakopoulos, "Towards Fully Autonomous Orbit Management for Low-Earth Orbit Satellites Based on Neuro-Evolutionary Algorithms and Deep Reinforcement Learning," *European Journal of Control*, vol. 80, p. 101052, Nov. 2024, doi: 10.1016/j.ejcon.2024.101052.

[11] K. Naik, O. Chang, and C. Kotulak, "Deep Reinforcement Learning for Autonomous Satellite Responsiveness to Observed Events," in *2024 IEEE Aerospace Conference*, Mar. 2024, pp. 1–10. doi: 10.1109/AERO58975.2024.10521008.

[12] H. H. Lei, M. Shubert, N. Damron, K. Lang, and S. Phillips, "Deep Reinforcement Learning for Multi-Agent Autonomous Satellite Inspection," in *Proceedings of the 44th Annual American Astronautical Society Guidance, Navigation, and Control Conference, 2022*, 2024, pp. 1391–1412. doi: 10.1007/978-3-031-51928-4_76.

[13] R. Guidotti, A. Monreale, S. Ruggieri, F. Turini, F. Giannotti, and D. Pedreschi, "A Survey of Methods for Explaining Black Box Models," *ACM Comput. Surv.*, vol. 51, no. 5, pp. 1–42, Aug. 2018, doi: 10.1145/3236009.

[14] M. Swartwout and K. J. Bennett, "DORRE: Autonomous Event Response Using an 8-Spacecraft Constellation," presented at the 38th Annual Small Satellite Conference, 2024. [Online].

Available: https://digitalcommons.usu.edu/smallsat/2024/all2024/33/

[15] Y. Li, "Deep Reinforcement Learning." Oct. 15, 2018. doi: 10.48550/arXiv.1810.06339.

[16] D. Quang Tran and S.-H. Bae, "Proximal Policy Optimization Through a Deep Reinforcement Learning Framework for Multiple Autonomous Vehicles at a Non-Signalized Intersection," *Applied Sciences*, vol. 10, no. 16, p. 5722, Jan. 2020, doi: 10.3390/app10165722.

[17] A. Herrmann and H. Schaub, "A Comparative Analysis of Reinforcement Learning Algorithms for Earth-Observing Satellite Scheduling," *Frontiers in Space Technologies*, vol. 4, Nov. 2023, doi: 10.3389/frspt.2023.1263489.

[18] J. Schulman, F. Wolski, P. Dhariwal, A. Radford, and O. Klimov, "Proximal Policy Optimization Algorithms." Aug. 28, 2017. doi: 10.48550/arXiv.1707.06347.

[19] M. A. Stephenson and H. Schaub, "BSK-RL: Modular, High-Fidelity Reinforcement Learning Environments for Spacecraft Tasking," in *75th International Astronautical Congress, Milan, Italy, IAF*, 2024, pp. 1186–1197. doi: 10.52202/078372-0120.

[20] P. W. Kenneally, S. Piggott, and H. Schaub, "Basilisk: A Flexible, Scalable and Modular Astrodynamics Simulation Framework," *Journal of Aerospace Information Systems*, vol. 17, no. 9, pp. 496–507, 2020, doi: 10.2514/1.I010762.

[21] P. Moritz *et al.*, "Ray: A Distributed Framework for Emerging AI Applications," presented at the 13th USENIX Symposium on Operating Systems Design and Implementation (OSDI 18), 2018, pp. 561–577. [Online]. Available: https://www.usenix.org/conference/osdi18/presentation/moritz

[22] "Rayproject/Ray - Docker Image | Docker Hub." https://hub.docker.com/r/rayproject/ray

[23] M. Stephenson, "Dynamics Sims — BSK-RL v1.1.16 Documentation." https://avslab.github.io/bsk_rl/api_reference/sim/dyn.html#properties

[24] H. Schaub, "Attitude Dynamics Fundamentals," *Encyclopedia of Aerospace Engineering*. Wiley, pp. 3181–3198, Dec. 15, 2010. doi: 10.1002/9780470686652.eae295.

[25] H. Schaub and J. L. Junkins, *Analytical Mechanics of Space Systems*, 2nd ed. Reston, Va: American Insti-

tute of Aeronautics and Astronautics, 2009. doi: 10.2514/4.867231.

[26] M. Stephenson, "FSW Sims — BSK-RL v1.1.16 1.1.16 Documentation." https://avslab.github.io/bsk_rl/api_reference/sim/fsw.html#actions

[27] N. J. Goldbaum, J. A. ZuHone, M. J. Turk, K. Kowalik, and A. L. Rosen, "Unyt: Handle, Manipulate, and Convert Data with Units in Python," *Journal of Open Source Software*, vol. 3, no. 28, p. 809, 2018, doi: 10.21105/joss.00809.